\documentclass[namedreferences,hyperref,optionalrh]{spr-sola}
\usepackage{graphicx}        
\usepackage[dvipsnames]{xcolor}           

\usepackage{gensymb} 

\newcommand{\p}[1]{{\color{blue}{#1}}}

\newcommand{\aap}{{\it Astron. Astrophys.}}

\newcommand{\apjl}{{\it Astrophys. J. Lett.}}
\newcommand{\apjs}{{\it Astrophys. J. Suppl.}}

\newcommand{\kms}{{km s$^{-1}$}}
\chardef\us=`\_

\begin{document}

\begin{frontmatter}
\title{Partial Null Point Reconnection of an Eruptive Filament\\
}

\author[addressref={aff1,aff2,aff3},corref,email={setiapooja.ps@gmail.com}]{\inits{P.}\fnm{Pooja}~\snm{Devi}\orcid{0000-0003-0713-0329}}

\author[addressref=aff4,email={mandrini@iafe.uba.ar}]{\inits{C. H.}\fnm{Cristina H. }~\snm{Mandrini}\orcid{0000-0001-9311-678X}}

\author[addressref=aff1,email={rchandra.ntl@gmail.com}]{\inits{R.}\fnm{Ramesh}~\snm{Chandra}\orcid{0000-0002-3518-5856}}

\author[addressref=aff4, email={gcristiani@gmail.com}]{\inits{G. D.}\fnm{Germán D.}~\snm{Cristiani}\orcid{0000-0003-1948-1548}}

\author[addressref=aff5, email= {Pascal.Demoulin@obspm.fr}]{\inits{P.}\fnm{Pascal}~\snm{D\'emoulin}\orcid{0000-0001-8215-6532}}

\author[addressref={aff6,aff7}, email= {cecilia.maccormack@nasa.gov}]{\inits{C.}\fnm{Cecilia}~\snm{Mac Cormack}\orcid{0009-0009-9799-979X}}

\author[addressref={aff6,aff8}, email= {dllovera@gmu.edu}]{\inits{D. G.}\fnm{Diego G.}~\snm{Lloveras}\orcid{0000-0003-1402-0398}}

\address[id=aff1]{Department of Physics, DSB Campus, Kumaun University, Nainital - 263 001, India}
\address[id=aff2]{Rosseland Centre for Solar Physics, University of Oslo, P.O. Box 1029, Blindern, N-0315 Oslo, Norway}
\address[id=aff3]{Institute of Theoretical Astrophysics, University of Oslo, P.O. Box 1029, Blindern, N-0315 Oslo, Norway}
\address[id=aff4]{Instituto de Astronomía y Física del Espacio, IAFE, UBA-CONICET, Pab. IAFE, Ciudad Universitaria, 1428 Buenos Aires, Argentina}
\address[id=aff5]{LIRA, Observatoire de Paris, Universit\'e PSL, CNRS, Sorbonne Universit\'e, Univ. Paris Diderot, Sorbonne Paris Cit\'e, 5 place Jules Janssen, 92195 Meudon, France}
\address[id=aff6]{Heliophysics Science Division, NASA Goddard Space Flight Center, 8800 Greenbelt Rd., Greenbelt, MD 20770, USA}
\address[id=aff7]{The Catholic University of America, Washington, DC 20064, USA}
\address[id=aff8]{Physics and Astronomy Department, George Mason University, Fairfax, VA 22030, USA}


\runningauthor{P. Devi et al.}
\runningtitle{Null Point Reconnection of an Eruptive Filament}

\begin{abstract}
Solar filaments are cool and dense plasma structures suspended in the solar corona against gravity. We present observations of a quiescent filament eruption that occurs on 13 July 2015. The eruption is associated with a two-ribbon GOES B8.9 class flare. Photospheric magnetic flux cancellation is present below the filament during days. This builds up a flux rope which progressively rises until it gets unstable, first leading to a confined eruption and pre-flare brightenings, then to an ejection which starts $\approx$ 20 min later with the flare onset. An interesting feature of this event is the presence of a large circular brightening formed around the erupting region.  This brightening is produced due to interchange reconnection of the ejected magnetic configuration with the surrounding open magnetic field. This null-point topology is confirmed by a potential-field extrapolation. The EUV loops located on the southern side of the filament eruption first contract during the null-point reconnection, then expand as the flux rope is ejected. The associated CME has both a classical flux rope shape and plasma ejected along open field lines on the flux rope side (a trace of interchange reconnection).
Finally, we set all this disparate observations within a coherent framework where magnetic reconnection occurs both below and above the erupting filament.
\end{abstract}
\end{frontmatter}

\section{Introduction}  
\label{sec:Introduction}

Solar filaments/prominences are dark/bright elongated features present in the corona. They are cooler ($\sim$ 10$^4$ K) and denser ($\sim$ 10$^9$ -- 10$^{11}$ cm$^{-3}$) by a factor $\approx$ 100 than the surrounding coronal atmosphere \citep[see the reviews by,][]{Labrosse2010, Mackay2010, Parenti2014}. They can be observed in active, as well as, quiet solar regions.  Usually, the magnetic configuration supporting filaments remains in a stable state  due to a balance between magnetic tension and magnetic pressure. Sometimes, the balance between magnetic tension and pressure is disturbed and, as a result, filaments are ejected outward in phenomena termed  as filament eruptions. Filaments can erupt fully or partially. The cases during which the plasma falls back to the chromosphere are known as failed eruptions \citep{Joshi2013, Cheng2015, Thalmann2015, Chandra2017twostep, Filippov2021, Joshi2022}. A partial or full filament eruption can evolve into a coronal mass ejection (CME) as observed in space- and ground-based coronagraphs.

Usually filament eruptions are associated with solar flares. Based on their morphology, flares can be classified as two-ribbon, multi-ribbon, and circular ribbon flares \citep{Parker1963, Svestka1986, Fletcher2011, Shibata2011, Wang2012, Chandra2017, Devi2020, Ibrahim2021}. In confined flares, the flare source is compact and such flares are not associated with CMEs. Two-ribbon flares are usually associated with CMEs and the ribbons separate as a function of time \citep{Svestka1986, Shibata2011, Benz2017}. Moreover, with the use of high-resolution observations, most flares are now observed as multi-ribbon flares \citep[for example see,][]{Ning2018, Joshi2021, Joshi2022, Dahlin2025, Faber2025}.

Anemone-like active regions (ARs) are those in which one polarity is surrounded by the opposite sign polarity \citep{Asai2008, Asai2009, Lugaz2011, Baker2013}. In such ARs, circular-ribbon flares  are observed \citep{Wang2012, Devi2020, Ibrahim2021}. 
Circular-ribbon flares may or may not be associated with CMEs \citep{Masson2009, Zuccarello2017, Hernandez-Perez2017, Hong2017, Li2018, Devi2020}.
In most circular-ribbon flares, a jet and a remote bright ribbon or kernel are also observed \citep{Masson2009, Pariat2010, Wang2012, Hernandez-Perez2017, Li2018surge, Lee2020, Mitra2021, Zhang2022} and the magnetic topology of the region is usually associated with the presence of a magnetic null point.

An interesting phenomenon is the change in the geometry of coronal loops or loop systems during filament eruptions and solar flares. These loop systems are different from flare loops and are present in the neighborhood of the eruption and flare source.
These loops expand/contract or oscillate during the eruptions. 
Observational studies as well as numerical simulations on the coronal and EUV loop dynamics accompanying filament eruptions 
and solar flares have been carried out \citep{Hudson2004, Mrozek2011, Dudik2017, Zuccarello2017, Chandra2021, Devi2021, Wang2021, Zhang2023}. 
\cite{Hudson2004} conduct a statistical study of 28 cases which show coronal loop oscillations. 
 \cite{Dudik2017} analyze two eruptive flares that are accompanied by the expansion/contraction of neighboring loops. These authors explain the loop dynamics as resulting from vortex formation, as proposed by \cite{Zuccarello2017}.
Another case study supporting vortex formation is presented by \cite{Devi2021}. In this study, the coronal loops begin to expand before the start of the filament eruption and then contract to regain their equilibrium position. 
\cite{Chandra2021} discuss a case study of coronal loops that differs in the sense that the loops start to contract after the start of the eruption. Moreover not only the loops present at the end of the erupting filament but also the ones over the stable part of the filament contract.

Despite the existence of many studies about filament eruptions and their relation to the dynamics of nearby loops, the origin of this dynamics is not clearly understood. With this in mind, we examine a filament eruption that occurred on 13 July 2015; the eruption is accompanied by a GOES B8.9 class flare and a CME. The filament erupted in three different phases, which is different to the classical two-phase eruption.  Furthermore, we observe the presence of a circular ribbon at a significant distance from the classical two flare ribbons. More surprising is the dynamics of the nearby loops which are first compressed and then they expand. Therefore, understanding the whole chain of events in this eruption is challenging. Indeed, because of the environment in which this eruption occurs, these observations provide a challenge to standard models and present numerical simulations of filament eruptions.

The paper is structured as follows. 
Section \ref{sec:Temporal_Sequence} outlines the temporal sequence of the event, covering the pre-flare and the impulsive and main flare phases, the presence of a circular brightening, loop expansion and contraction, and the associated CME. This section also details the evolution of the photospheric magnetic field before, during, and after the eruption. Section \ref{sec:Physical_Processes} provides an interpretation of the underlying physical processes. Finally, in Section \ref{sec:Conclusion} we summarize the results and conclude. The time-line for the event of 13 July 2015 is given in Table \ref{tab:chronology}. Details of the instruments and data sets used in this study are provided in Appendix \ref{sec:obs}.

\setlength{\tabcolsep}{2pt}
\renewcommand{\arraystretch}{1.3}
\begin{table}
\caption{Table for the chronology of the event on 13 July 2015.} 
    \centering
    \begin{tabular}{ll}
    \hline
Observed activity&  Time or duration\\
\hline
Brightening below the filament in AIA 304 and 171 \AA\ & 8:48 UT \\
Onset of pre-flare or flare precursor in GOES & 8:50 UT \\
(time close to the brightening visible in AIA 304 and 171 \AA) & \\
Duration of the pre-flare phase& 8:50 - 9:10 UT \\
Start of filament rise & 8:50 UT \\
Contraction of a system of loops at the south end of the filament & 8:52 UT - 9:12 UT \\
GOES B8.9 class flare onset & 9:10 UT \\
Two reverse  J-shaped flare ribbons clearly visible & 9:10 UT \\
Brightening at [-680$''$, -180$''$] at the south of the filament & 9:10 UT \\
Loop-system expansion & 9:12 UT - 9:50 UT \\
Full circular ribbon appearance  & 9:22 UT  \\
(continued for a long time, more than an hour) & \\
Flare loops visible in AIA 171 \AA\ & 9:28 UT \\
GOES B8.9 class flare peak & 9:30 UT \\
\hline
    \end{tabular}
    \label{tab:chronology}
\end{table}

\section{Temporal Sequence of the Event} 
\label{sec:Temporal_Sequence}
The event on 13 April 2015 is accompanied by a filament eruption, a two-ribbon flare, the formation of a circular brightening around the eruption region, the contraction and expansion of a neighboring system of loops, and a CME. The results at different atmospheric heights are presented in the following sub-sections.

\begin{figure}[!t]
    \centering
    \includegraphics[width=\textwidth]{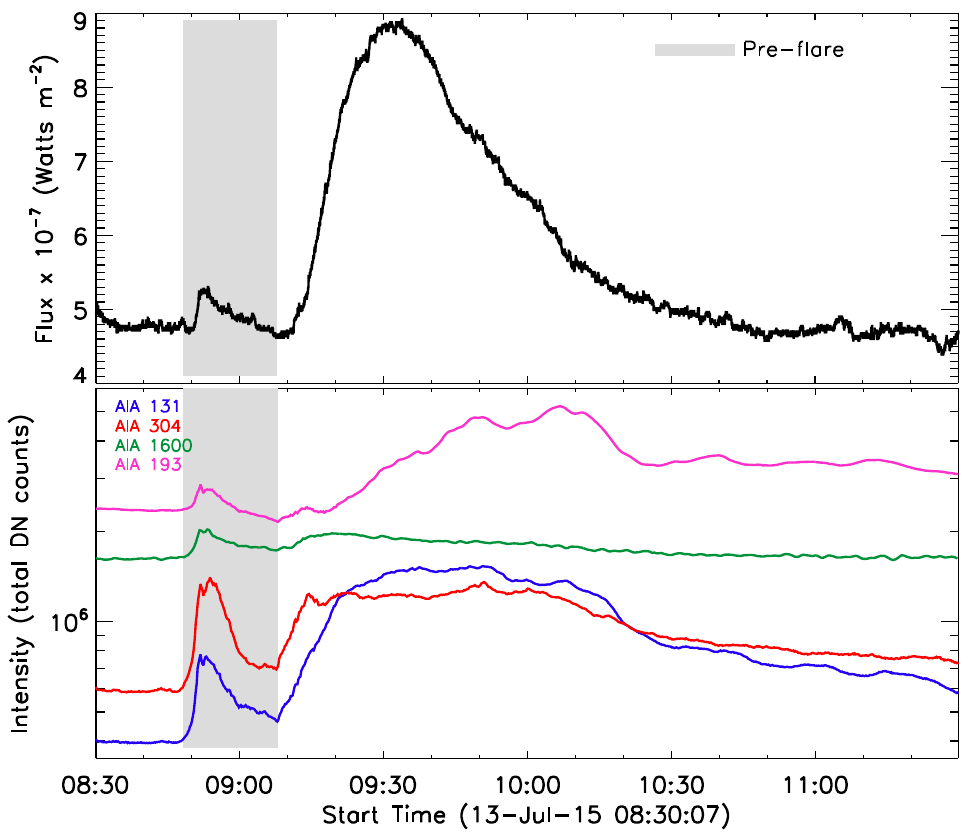}
    \caption{Top panel: Temporal evolution of GOES X-ray flux in the  1 -- 8 \AA\ wavelength range. 
    Bottom panel: Temporal variation of the intensity of AIA in different wavelengths accumulated in the black box shown in Figure \ref{fig:evolution}h. For better visualization, the intensity of AIA 193 \AA\ is divided by 6.
    Plots with different colors correspond to different wavelengths as indicated on top left corner of the panel. The gray shaded region highlights the pre-flare brightening in the active region. }
    \label{fig:goes_aia}
\end{figure}

\subsection{Pre-flare, Impulsive and Main Phases of the B8.9 Class Flare}
\label{sec:flare}

The top panel of Figure \ref{fig:goes_aia} shows a pre-flare enhancement in the GOES light curve before the flare. This pre-flare enhancement begins at $\approx$ 8:48 UT, peaks around 8:52 UT, and then decays until $\approx$ 9:10 UT. The duration of the pre-flare is highlighted with a gray background in Figure \ref{fig:goes_aia}. After 9:10 UT, the flux in the GOES 1–8 \AA\ waveband  starts increasing again and reaches its peak at $\approx$ 9:30 UT.

\begin{figure}[!t]
    \centering
    \includegraphics[width=\textwidth]{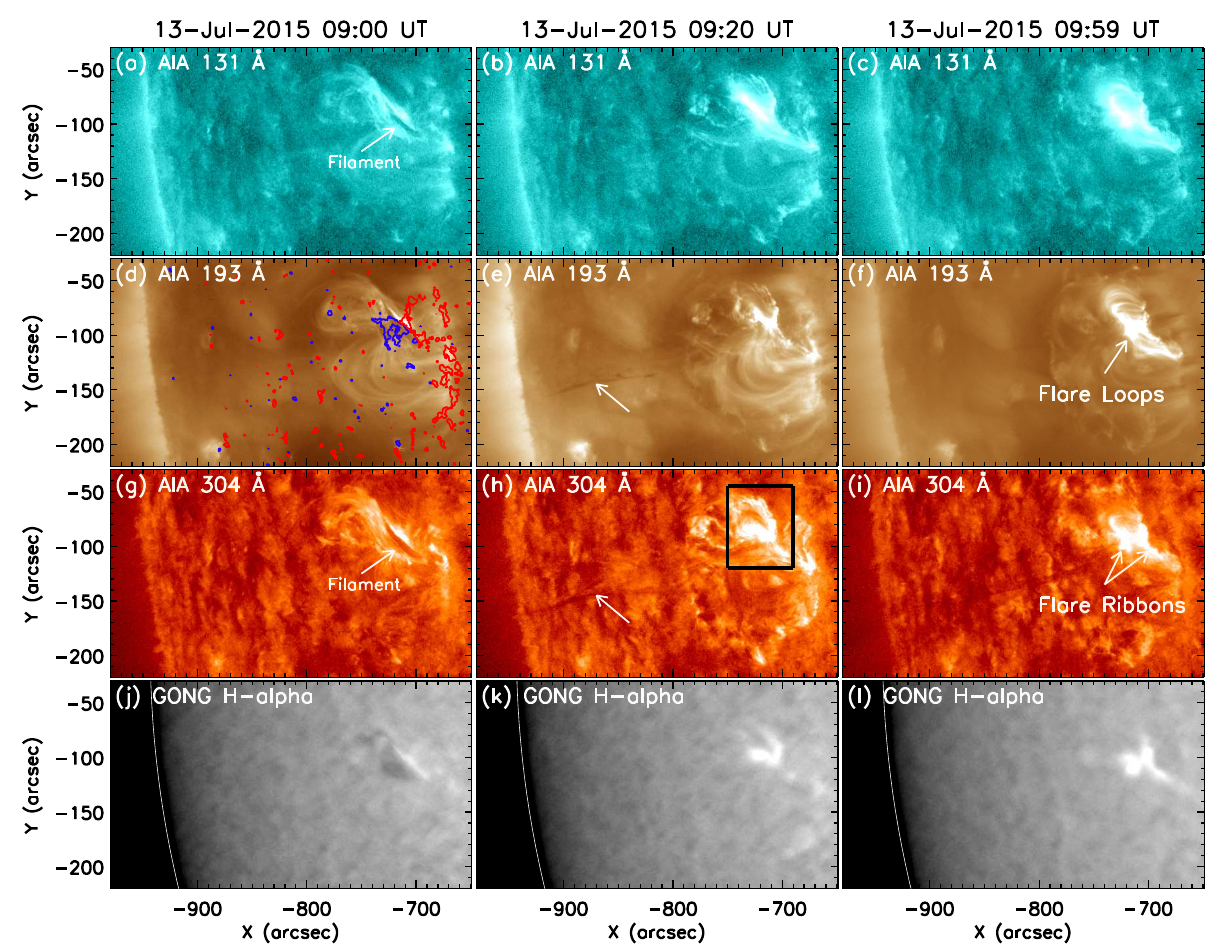}
    \caption{Evolution of the filament eruption in AIA 131 \AA\ (a -- c), 193 \AA\ (d -- f), 304 \AA\ (g -- i), and in GONG H$\alpha$ (j -- l) wavelengths. The red and blue contours in panel d are the contours of positive and negative magnetic polarities ($\pm$ 40 G) from HMI, respectively. The black box marks the region used to calculate the AIA light curves in Figure~\ref{fig:goes_aia} (bottom panel). The erupting filament is shown with white arrows in panels e, g, and h at different projected heights. The flare loops and flare ribbons are pointed with arrows in panels f and i. An animation (MOV\_Fig2.mp4), related to this figure, is included as online supplementary material.} 
    \label{fig:evolution}
\end{figure}

We also plot the intensity light curves of different AIA wavelengths in the filament eruption region. This region is shown with a black box in Figure \ref{fig:evolution}h. These light curves are shown in the bottom panel of Figure \ref{fig:goes_aia}. The pre-flare enhancements in AIA starts at the same time as in GOES and lasts for $\approx$ 20 min. 
As shown in the movie attached to Figure \ref{fig:evolution}, the filament goes mostly out of the selected box after 9:10 UT; so its contribution to the intensity in Figure \ref{fig:goes_aia} is limited to the pre-flare phase. During this early phase, it is mostly absorbing the EUV radiation located behind it along the line of sight, and also plausibly contributing to extra EUV emission because part of its dense plasma is heated up.

Similarly to the GOES curve, the flare  begins at $\approx$ 9:10 UT in almost all AIA wavelengths.  As shown in the bottom panel of Figure \ref{fig:goes_aia}, a small intensity bump is present in AIA 193 \AA\ during the sudden intensity increase of the flare, i.e. the impulsive phase. This small bump appears at $\approx$ 9:15 UT. The intensity variation with time in AIA 171 and 211 \AA\ is similar to AIA 193 \AA. Therefore, to avoid overlapping the plots, we do not show the curves of AIA 171 and 211 \AA. The peak of the flare in AIA 1600 \AA\ appears before the peak in GOES 1 -- 8 \AA\ and the intensity evolution has a longer decay. In addition, in AIA 304 and 131 \AA\ intensity plots the higher intensity period lasts longer, $\approx$ 50 min, from $\approx$ 9:15 -- 10:05 UT. 
On the other hand, the light curves of AIA 171, 193, and 211 \AA\ show a global enhancement delayed by $\approx$ 15 min (at $\approx$ 9:25 UT) compared to the other wavelengths. 

\begin{figure}[!t]
    \centering
    \includegraphics[width=0.9\textwidth]{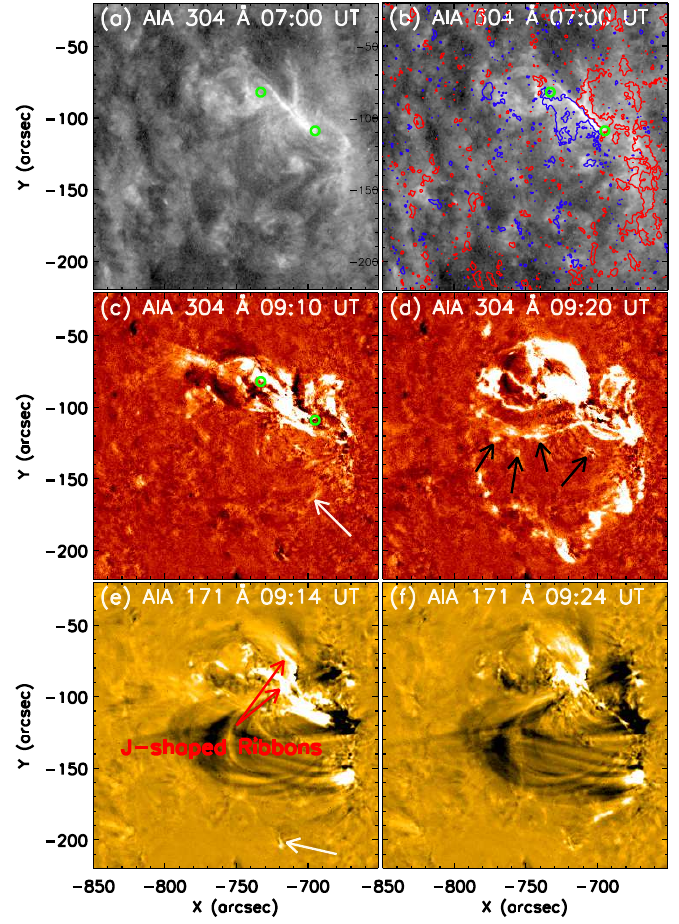}
    \caption{Images of AIA 304 \AA\ in the top panels show the filament feet with green circles. Panel b is the same as panel a with magnetic field contours ($\pm$ 20 G) from SDO/HMI, red for the positive and blue for negative magnetic polarity. The middle and bottom panels are the base difference images (subtracted from the image at 08:30 UT) showing the large circular ribbon around the erupting filament. The initial circular ribbon brightening is shown in (c) and (e) with white arrows. The black arrows in (d) show the inner brightenings inside the circular one. Red arrows in (d) show two inverse J-shaped flare ribbons. 
    }
    \label{fig:circular_ribbon}
\end{figure}

\subsection{Filament Eruption}
\label{sec:Filament_Eruption}

The filament was located at S03 E50 on 13 April 2015. The evolution of the eruption in AIA 131, 193, 304 \AA\ and in GONG H$\alpha$ is presented in Figure \ref{fig:evolution}. Before the eruption, the filament is shown by an arrow in panels a and g. The contours of HMI line-of-sight (LOS) magnetic field at $\pm$ 40 G are over-plotted on the AIA 193 \AA\ image in panel d. The overlay shows that the filament is located between two opposite magnetic polarities along the polarity inversion line (PIL).

From the movie associated with Figure \ref{fig:evolution} (MOV\_Fig2.mp4), a brightening is present below the filament at $\approx$ 8:48 UT, this corresponds to the pre-flare enhancement seen in the GOES and AIA intensity curves of Figure \ref{fig:goes_aia}.
This brightening intensifies and extends along the full filament at $\approx$ 8:50 UT. The filament seems to be destabilized during the pre-flare phase; however, it does not erupt as it  is still observed along the PIL at 9:00 UT (see, mainly, panels a and g of Figure \ref{fig:evolution}).

The filament erupted during the impulsive phase of the B8.9 class flare. The erupting filament is shown in panels e and h of Figure \ref{fig:evolution}. During the eruption of the filament, a left-handed twist is clearly visible in almost all the AIA EUV wavelengths (see the accompanying movie, MOV\_Fig2.mp4). The eruption is associated with the presence of two inverse J-shaped flare ribbons that are seen in all AIA and in H$\alpha$ wavelengths (panels b, e, h, and k). 
These ribbons separate from each other as a function of time, as seen in typical two-ribbon flares. After the peak phase of the flare, post-flare loops are connecting the two inverse J-shaped ribbons. These loops appear sheared, as can be seen in panel f (compared to the PIL orientation in panel d).

As the filament reaches higher heights (projected height about 40 Mm), a circular brightening is observed surrounding the two flare ribbons (panel h of Figure \ref{fig:evolution}). To see the evolution of the circular ribbon more clearly, we  use the base difference technique. For this, we subtract the image at 8:30 UT from the other images. The base difference images in AIA 304 and 171 \AA\ are displayed in Figure \ref{fig:circular_ribbon} middle and bottom panels, respectively. The circular brightening appears around $\approx$ 9:10 UT in AIA 304 \AA\ and is visible at $\approx$ [-680$''$, -180$''$], i.e., southward of the erupting region. However, in the case of AIA 171 \AA, this brightening appears $\approx$ 2 min later at $\approx$ [-720$''$,-205$''$]. Probably, because it took about 2 min to heat up the plasma higher in the corona to be seen in 171 \AA.
This brightening extends with time to form a large circular emission region mostly southward of the two inverse J-shaped ribbons. The circular brightening formed more than half a circle by $\approx$ 9:20 UT. After its formation, it remained visible for a long time (more than an hour). In addition to the large circular brightening, inner brightenings are also present inside the circular one. The inner brightenings are shown by black arrows in Figure \ref{fig:circular_ribbon}d.
The circular and inner brightenings may occur because of the interaction of the erupting flux rope with neighboring high-lying/open magnetic field lines (see Section \ref{sec:PFSS}).
Figures \ref{fig:circular_ribbon}a and \ref{fig:circular_ribbon}b present the images of AIA 304 \AA\ before the eruption, with and without the overlaid contours of the magnetic field, respectively. 
Careful analysis of the data reveals that the northern footpoint of the filament is rooted in the negative magnetic polarity, while the southern footpoint lies in the positive polarity. This observation aligns with the magnetic environment near the polarity inversion line (PIL) and the magnetic helicity of the local magnetic field \citep[e.g.,][]{Amari2003}.

\begin{figure}[!t]
    \centering
    \includegraphics[width=\textwidth]{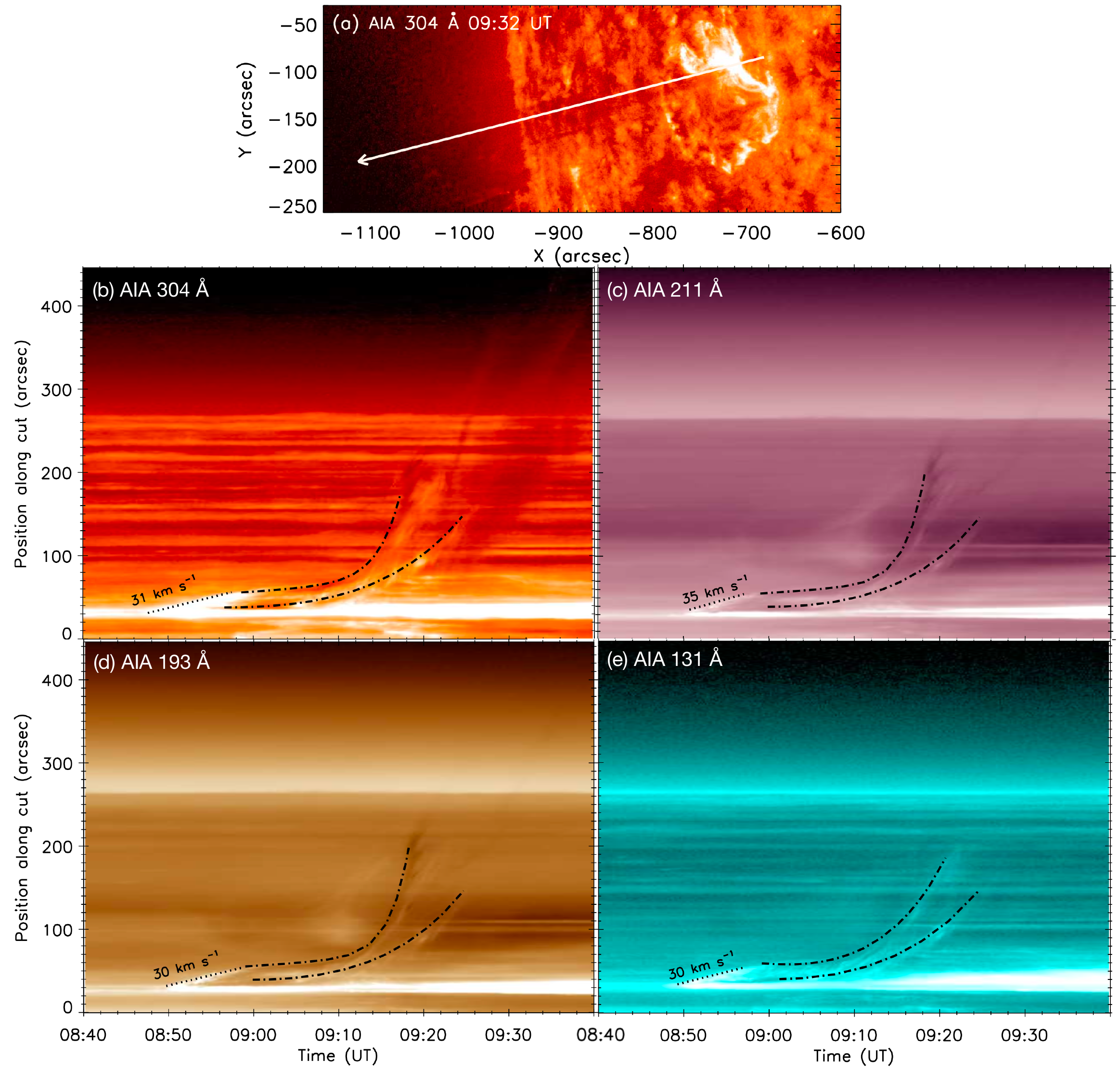}
    \caption{Time-distance analysis of the filament eruption in different AIA wavelengths. (a): AIA image in 304 \AA\ showing the selected direction with a white arrow. (b), (c), (d), and (e): Display of the time-distance diagram along the chosen direction in AIA 304, 211, 193, and 131 \AA, respectively. 
    The dotted and dash-dotted lines in the time-distance diagrams are the fittings of linear and a combination of linear and exponential functions, respectively.}
    \label{fig:kinematics}
\end{figure}

To analyze the kinematics of the filament eruption, we compute time-distance plots. For this analysis, we select a slice in the initial direction of the filament eruption, which is shown by a white arrow in Figure \ref{fig:kinematics}a. The time-distance diagrams in different AIA wavelength bands are shown in panels b, c, d, and e. A bright structure, the heated filament, is observed rising. Then we infer that the filament starts to rise at $\approx$ 8:48 UT. This is the time when the brightenings are observed during the pre-flare phase in GOES and AIA light curves (see Figure~\ref{fig:goes_aia}). We trace the leading part of the moving filament to get its heights at different times. These data are fitted between 8:48 and 8:58 UT with a straight line shown by dotted lines in Figure \ref{fig:kinematics}b -- e. The speeds computed from this fitting in each wavelength are written over the dotted lines in each panel. The average speed is found to be $\approx$ 32.5 $\pm$ 2.5 \kms.
From the time-distance diagrams, the filament rises up to a projected height of $\approx$ 30$''$, i.e., $\approx$ 22 Mm at 8:58 UT. After reaching this height, the filament nearly stops rising for a while. 

About 10 min later the filament erupts again. Then, starting
by around 9:00 UT, we fit the position-time data, $s(t)$, with a combination of linear and exponential functions, given by 
  \begin{equation} \label{Eq-fit}
    s = ae^{bt}+ct+h_0
  \end{equation}
where, $a,~b,~c,$ and $h_0$ are constants to be found by the least square fitting. For this fitting, we use the $mpfit$ function in Solarsoft (SSW). The results are overlayed on the time-distance plots with dash-dotted lines. 
The value of $c$ gives the speed of the linear motion of the erupting filament corresponding to 
the slow rise-phase of the filament, when the first term in Equation \ref{Eq-fit} is negligible. 
The average speed of this slow rise-phase varies from 6 to 9 \kms\ in different AIA wavelengths, so significantly lower than $\sim 32$ \kms\ found in the first phase.

The acceleration phase starts when the exponential part of the above mentioned fitting function dominates over the linear part. This is seen after the slow linear rise of the filament in Figure \ref{fig:kinematics}b -- e. It is noticeable from the time-distance diagrams and movies related to Figure \ref{fig:evolution} that the acceleration phase of the eruption starts at $\approx$ 9:10 UT. This corresponds approximately to the flare start (Figure \ref{fig:goes_aia}). The average speed of this phase varies from $\approx$ 140 -- 200 \kms. 

The speed range in the slow and acceleration phases, as fitted with Equation \ref{Eq-fit}, is consistent with previous observations and models that conclude that filaments erupt in two different phases: the slow rising phase and the acceleration phase \citep[for example,][]{Schrijver2008, Cheng2023, Xing2024}. 
For example, \cite{Schrijver2008} study two filament eruptions and find that the filaments exhibit a speed of $\approx$ 4 and 13 \kms\ during their slow rising phase and then the filaments erupt with a much higher speed. However, here in contrast to previous studies, the evolution is more complex. The filament begins to rise with $\approx$ 30 \kms, then slows down to 6 -- 9 \kms, and later accelerates exponentially, reaching a speed of up to 200 \kms.

From the time-distance diagrams, we see another strand of the filament following the main filament. This strand is visible in almost all AIA wavelengths at $\approx$ 40$''$ distance at 08:54 UT when it is first visible (this is clearer in panel e, AIA 131 \AA , as a dark feature). Then, this distance increases progressively with time. The position-time data are fitted using Equation \ref{Eq-fit}. 
This fit is shown by the lower dash-dotted line in Figure \ref{fig:kinematics}b -- e. 
For this strand, the speed of the linear part is found to vary from $\approx$ 13 to 22 \kms, i.e. about a factor of 2 faster than that of the leading edge of the filament. However, later on the speed during the acceleration phase is slower. 
The average speed of the exponential rise ranges from $\approx$ 70 to 80 \kms\ in different AIA wavelengths, i.e. about a factor of 2 slower than the leading edge.
These results possibly indicate that the instability starts at a lower height, where we find  the faster speed in the linear phase. Later on, magnetic reconnection above the erupting filament (see Section \ref{sec:Physical_Processes}) removes part of the filament overlying arcade field lines, implying a decrease of  the downward magnetic tension and allowing a faster rise of the leading edge.

\label{sec:loops}
\begin{figure}[!t]
    \centering
    \includegraphics[width=\textwidth]{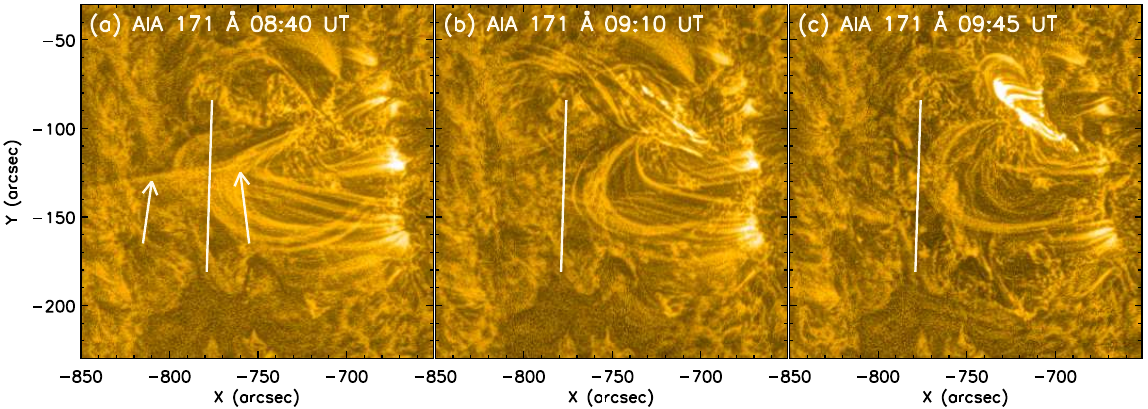}
    \caption{Images of AIA 171 \AA, using the MGN method, showing the evolution of the EUV loops at the south of the erupting filament. The white vertical lines represent the top of one set of loops before the eruption, i.e. at 8:40 UT. In panel a, two white arrows point to : (left) a straight elongated shape suggesting the presence of open magnetic field lines and (right) very stretched loops. The same vertical line has been added to panels b and c. An animation (MOV\_Fig5.mp4) of this figure is available.
    }
    \label{fig:loops}
\end{figure}

\begin{figure}[!t]
    \centering
    \includegraphics[width=0.95\textwidth]{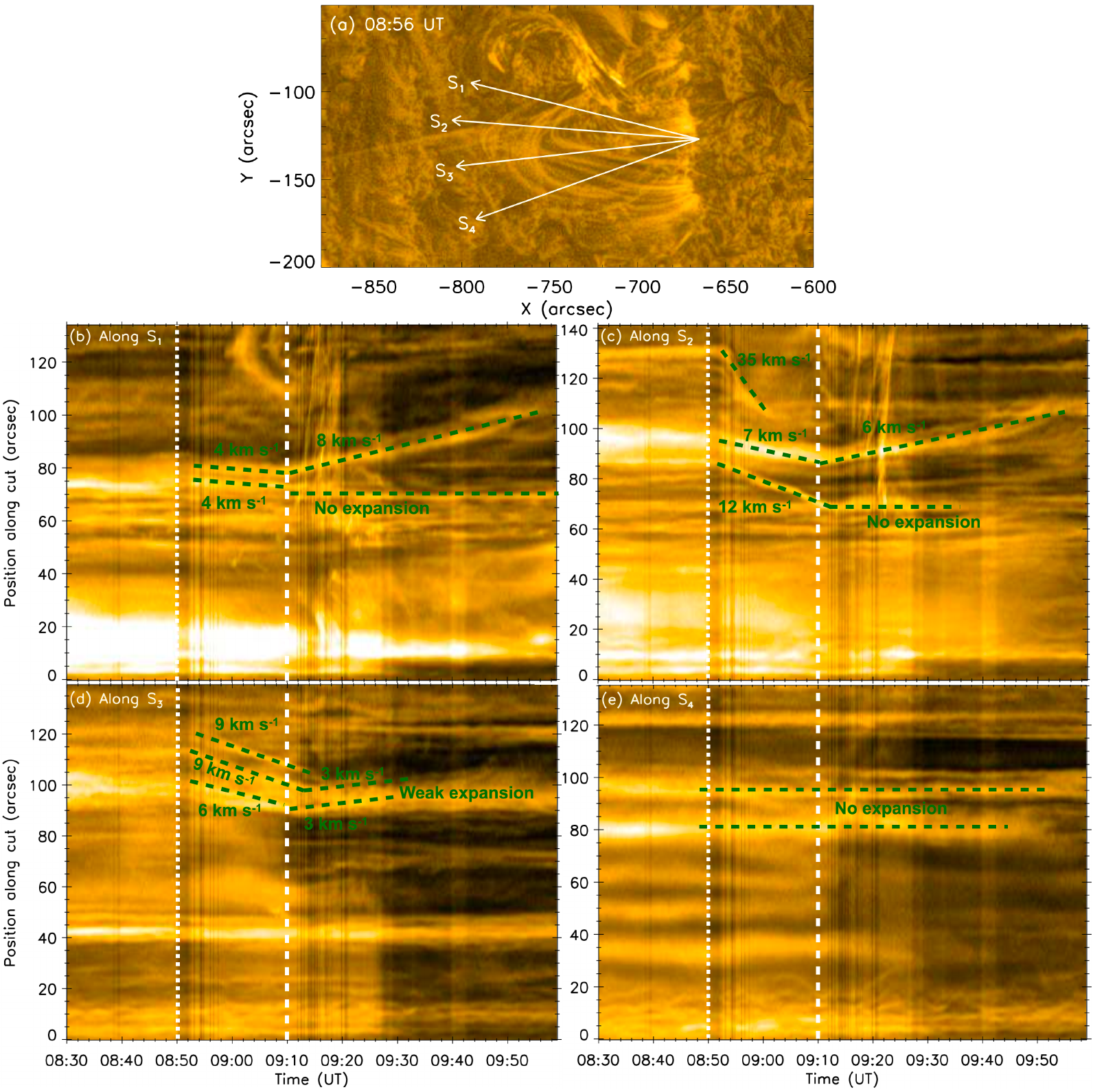}
    \caption{Time distance analysis of the loops located southern to the erupting filament. (a):  The directions S$_1$, S$_2$, S$_3$, and S$_4$ along the EUV loops in AIA 171 \AA\ images. The time-distance plots along these directions are displayed in panels b -- e. The white dotted and dashed lines represent the start time of the pre-flare enhancement and flare, respectively. The MGN technique is applied to the images to enhance the EUV loops. }
    \label{fig:ts_loop}
\end{figure}

\subsection{EUV Loop Dynamics}
\label{sec:EUV_Loop}

We study the change in the EUV loop system present southward of the filament. For this purpose, we have used the Multi-Gaussian Normalization (MGN) method. 
This method improves the visibility of EUV loops by normalizing every pixel of the images using the method explained by \cite{Morgan2014}. This method has been used by several authors in previous studies \citep[for example;][]{Chandra2021, Devi2021, Li2023, Luna2024}. The resulting MGN images are shown in Figure \ref{fig:loops}.

An interesting feature shown in Figure \ref{fig:loops}a is the presence of a ``cusp-like shape'' above the loops (in the region between the white arrow heads).   This cusp-like shape suggests the possible presence of a magnetic null point. From the observations, we can get a crude estimate of the height of this null point. First, we estimate the distance between the projected location of the cusp-like shape and the nearest point of the PIL.
Then, the null point height is computed as the ratio between this distance and the sine(longitude from central meridian), supposing that the null is approximately above the PIL and neglecting the latitude contribution. The estimated distance is $\approx$ 55$''$, so $\approx$ 40 Mm, and the longitude at which the PIL is located is $\approx$ 45$\degree$. Then, the height of this probable null point is $\approx$ 56 Mm. 

Next, the evolution of the EUV loops in AIA 171 \AA\ using the  MGN method is shown in Figure \ref{fig:loops}. 
The north-south white line is set at the top of the loop system before the loops start to contract (panel a). In panel b, the loops appear shifted toward the west  of the same white  line. 
Taking into account that the region is located on the eastern side of the solar disk, this means that the loops contract. This continues until $\approx$ 9:10 UT (see Figure \ref{fig:ts_loop}), i.e. until the end of the pre-flare enhancement in GOES and AIA wavelengths. 
After 9:10 UT, the EUV loops start to expand (see the animation MOV\_Fig5.mp4 related to Figure \ref{fig:loops}). 
This expansion is accompanied by the appearance of the circular brightening shown in Figures \ref{fig:evolution}h and \ref{fig:circular_ribbon}. 

To study the dynamics of the EUV loops, we perform a time-distance analysis. For this analysis, we select slices along four different directions S$_1$, S$_2$, S$_3$, and S$_4$ as shown in panel a of Figure \ref{fig:ts_loop}. The corresponding time-distance diagrams are shown in panels b, c, d, and e, respectively. The white dotted and dashed vertical lines in these plots correspond to the start of the pre-flare and of the main phase of the B8.9 class flare. The dynamics of the loops, projected along the slits, is as follows: 
\begin{enumerate}
    \item {\bf Along S$_1$:}
S$_1$ is chosen close to the erupting filament. During the pre-flare enhancement, which starts at $\approx$ 8:50 UT, the loops contract with a speed of 4 \kms. This contraction  stops around 9:10 UT, which is the start of the flare. After this time, the loop closer to the slit origin, so at lower height and closer to the PIL, remains stable without much change in height. 
On the other hand, the higher loop expands with a speed double from the contraction speed, i.e. 8 \kms. This loop even reaches $\approx$ 29 Mm farther away from its original position. 
    \item {\bf Along S$_2$:}
S$_2$ covers three sets of loops as seen in its time-distance plot, panel c. Similar to S$_1$, all the loops begin to contract with the start of the pre-flare enhancement. The highest loops contract with the highest speed, which is $\approx$ 35 \kms\ and then disappear after $\approx$ 9:00 UT. 
The middle loop shows a contraction until the start of the flare, i.e. 9:10 UT with a speed of $\approx$ 7 \kms, and then it expands with a slightly lower speed ($\approx$ 6 \kms). This is the same as for S$_1$, and it reaches a  larger height than its initial one. The lower loop also contracts at same time (8:50 UT) but with approximately a double speed, i.e. 12 \kms\, and after 9:10 UT it reaches a stable position. 

   \item {\bf Along S$_3$:}
S$_3$ also covers the same three loops as direction S$_2$. However, the dynamics of the loops changes along this direction. The highest loop  contracts with a speed of 9 \kms\ until $\approx$ 9:14 UT and then disappears. 
The middle loop contracts with the same speed during the same time interval and then expands with a much lower speed of 3 \kms. 
This evolution is also seen in the lower loop, which shows contraction with a speed of $\approx$ 6 \kms\ and expansion with 3 \kms.  Another loop, which is present at a slit position of $\approx$ 40$''$ (so westward from the previous ones, panels a and d), does not show any contraction or expansion during the pre-flare and flare. Its top is indeed crossed by S$_2$ at a slit position of $\approx$ 60$''$, and it shows no evolution as well (Figure \ref{fig:ts_loop}, panel c). 

   \item {\bf Along S$_4$:}
S$_4$ is farther from the erupting filament region than the other three directions. From the time-distance plot along S$_4$, the eruption does not affect the loops at this location.
 \end{enumerate}

From all these time-distance plots, globally we conclude that the expansion speed of the same loops decreases when the chosen direction is farther from the eruption region or at a  height well below the estimated null point height. This contrasts with the contraction speed which is maximum near the loop apex. 

\begin{figure}[!t]
    \centering
    \includegraphics[width=\textwidth]{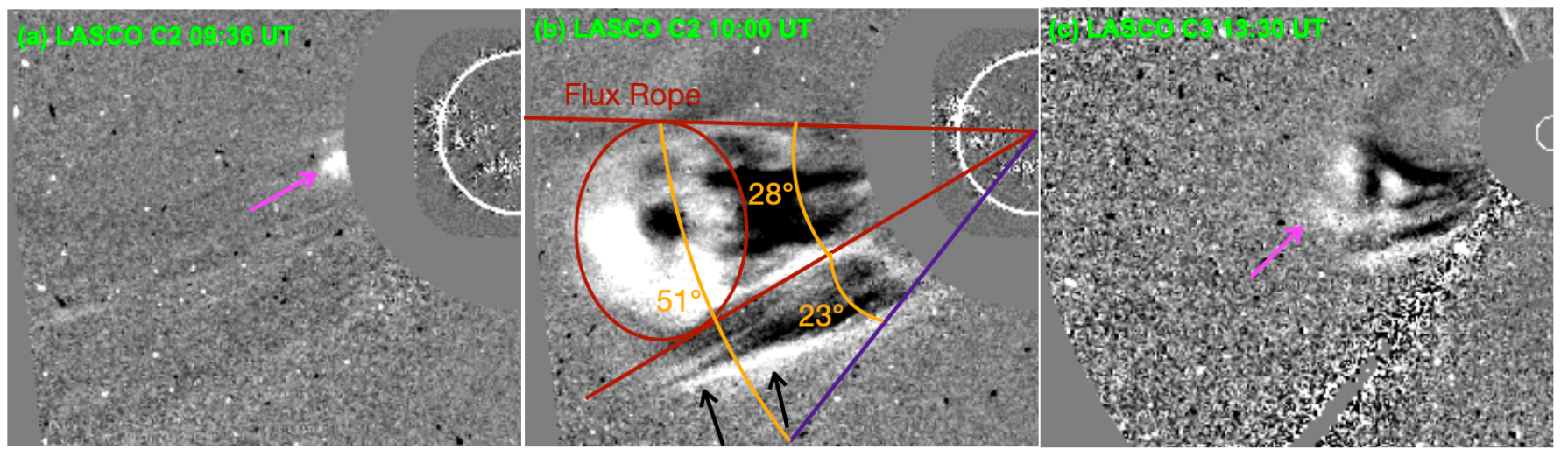}
    \caption{Evolution of the CME associated to the filament eruption observed in LASCO C2 and C3 coronagraphs. The CME leading edge is shown by pink arrows in panels a and c. Panel b shows the angular width of the flux rope (in brown color) and of the southward CME limit (purple). The angular extensions of these structures are indicated in orange. The black arrows mark the plasma ejected along open magnetic field lines.}
    \label{fig:CME}
\end{figure}

\subsection{Associated CME} \label{sec:CME}
The filament eruption on 13 July 2015 is accompanied by a CME. The evolution of the CME as seen by the Large Angle and Spectrometric Coronagraphs  (LASCO) C2 and C3 is shown in Figure \ref{fig:CME} using running difference images. The CME is visible in LASCO C2 at 9:36 UT at a height of 2.65 R$_\odot$ (panel a). According to the LASCO CME catalog (\url{https://cdaw.gsfc.nasa.gov/CME_list/}), the speed and acceleration of the CME are $\approx$ 553 \kms\ and $\approx$ -14 m s$^{-2}$, respectively. 

Two main features are visible in LASCO images.
First, we observe the ejection of a flux-rope like structure (Figure \ref{fig:CME}b). 
Second, part of the coronal plasma is ejected along open magnetic field lines present at the south of this flux rope structure, as shown by black arrows in panel b. 
We measure a full angular width of 51$\degree$ for the CME consisting of the flux rope, 28$\degree$ broad, and of the plasma ejected along the southward open field lines, 23$\degree$ broad. These estimations are not corrected because of projection effects, which broaden the estimated widths.
Our value of the CME width does not agree with that reported in LASCO catalog (104$\degree$), which does not seem to be realistic in view of the images shown in Figure \ref{fig:CME}.

\begin{figure}[!t]
    \centering
    \includegraphics[width=\textwidth]{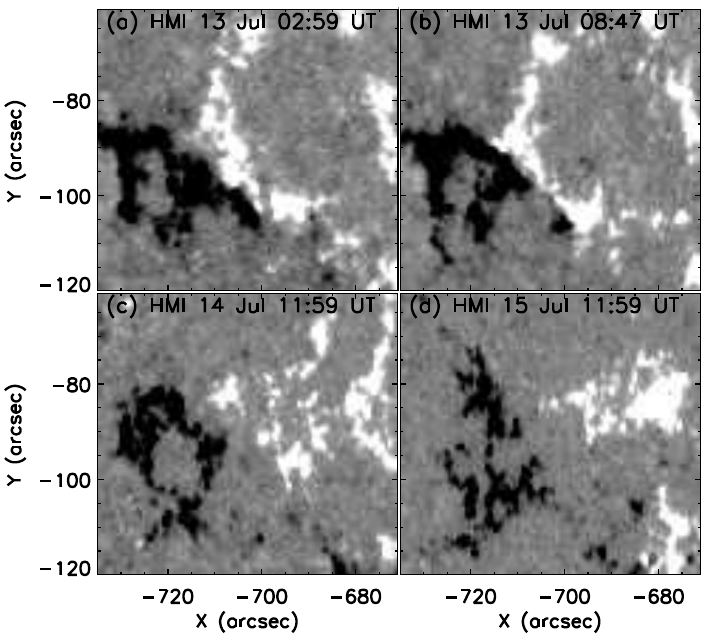}
    \caption{HMI LOS images showing the evolution of the photospheric magnetic field from 13 to 15 July 2015. A clear dispersion of the magnetic field concentrations is observed as the studied bipolar region evolves. The magnetic field values are saturated above/below $\pm$ 80 G. The HMI images are aligned at 08:30 UT on 13 July 2015.
    This figure is accompanied by a movie (MOV\_Fig8.mp4) showing the field evolution along 13 July covering most of the pre-flare phase and up to 15 July 2015 showing the evolution of the magnetic polarities even after the eruption.
    }
    \label{fig:evolution_hmi}
\end{figure}

\subsection{The Photospheric Magnetic Field}  \label{sec:Photospheric_B}

The studied region appears at the eastern limb on 11 July 2015 at around S03.  As this region never develops sunspots, it is not classified with a NOAA number. In this section we analyze the evolution of the photospheric magnetic field and its flux along a few days starting on the day of the eruption.

Figure \ref{fig:evolution_hmi} displays the evolution of the studied region in HMI LOS magnetograms.  The region is mainly bipolar with a PIL oriented in the NE-SW direction.  On 13 July 2015, the erupting filament  lies on this PIL. The polarities evolve and diffuse, as shown in panels c and d. By around 16 July only the negative polarity of the bipole can be identified, the positive polarity is largely dispersed due to the convective buffeting. 

\begin{figure}[!t]
    \centering
\includegraphics[width=0.9\textwidth]{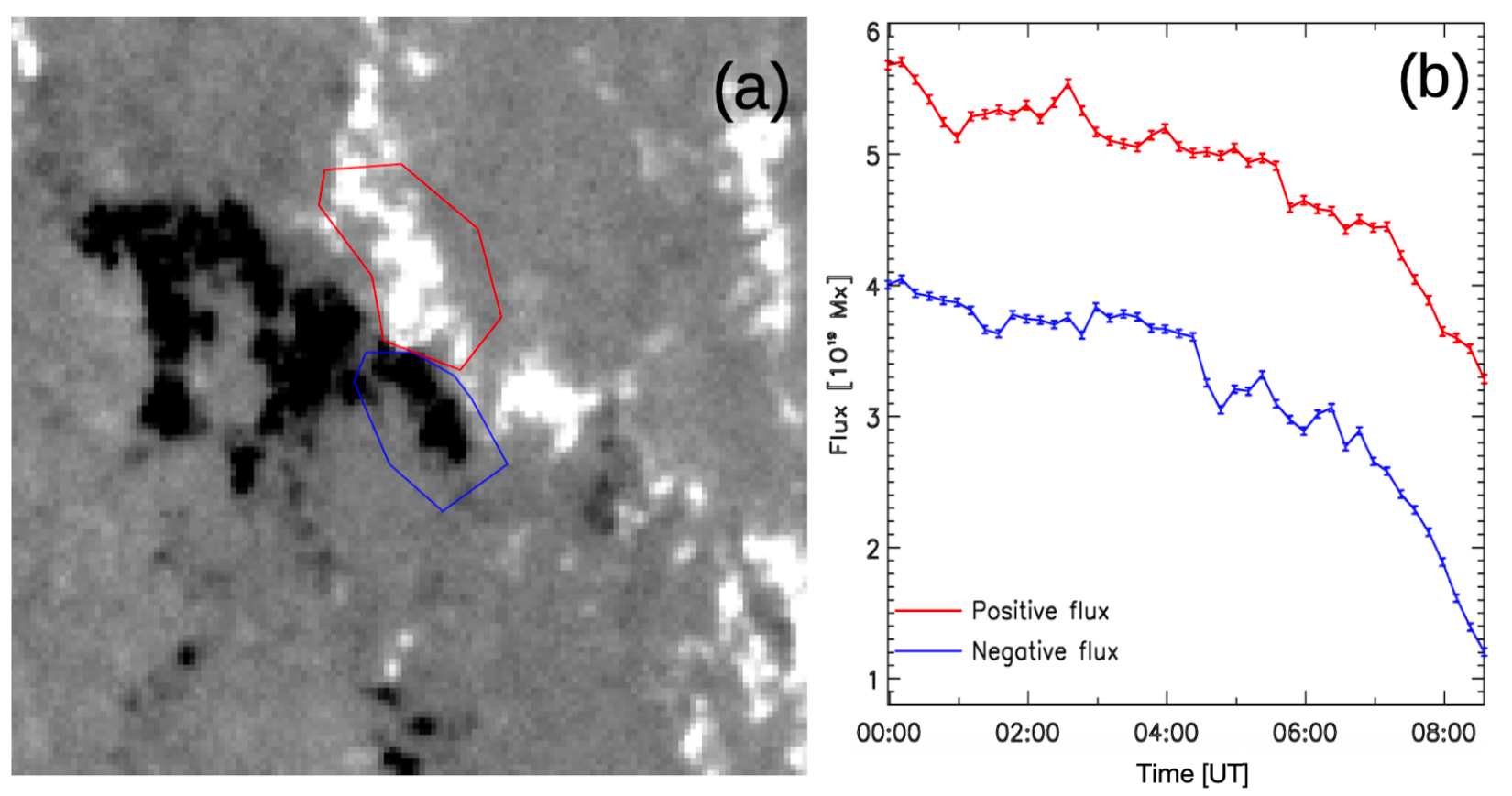}
    \caption{
    Computation of positive and negative magnetic flux at specific locations on both sides of the PIL. The locations of flux computations for positive and negative flux are shown in panel a surrounded by a red and blue curve, respectively. The magnetic field is saturated as in Figure \ref{fig:evolution_hmi}. The computed fluxes are plotted in panel b. The error bars in these curves are calculated considering that the magnetic field measurement error per pixel is of $\approx$ 5 G following a Poisson statistics. 
    }
    \label{fig:flux}
\end{figure}

In order to find clues for the filament destabilization and eruption, we compute the evolution of photospheric magnetic flux along the PIL. Magnetic flux cancellation is present at different sites along the PIL where the filament  is located. A clear site of flux cancellation is shown in Figure \ref{fig:flux}a on a magnetogram taken just before the start of pre-flare enhancement (at $\approx$ 8:40 UT). In this panel, the sites where we compute the evolution of the positive and negative magnetic fluxes are surrounded with red and blue curves, respectively. The estimated positive and negative flux in these regions is presented in panel b. There is a continuous decrease, with similar magnitude, of positive and negative magnetic fluxes with time. The cancellation in both positive and negative magnetic field zones  begins around 04:00 UT. The fluxes continue to decrease until 9:00 UT as shown in the plot. However, the decrease in flux goes on even later (not presented in the plot), as can be implied from Figure \ref{fig:evolution_hmi}c -- d. 

This flux cancellation could first contribute to build the flux rope by forcing the progressive transformation of the sheared arcade, present above the PIL, to twisted field lines as proposed by \citet{vanBallegooijen1989,vanBallegooijen1990}. The build up of this flux rope creates a favorable configuration with magnetic dips to support dense and cold plasma. During magnetic reconnection dense plasma is expected to be lifted up in the flux rope. Then, the filament is expected to be embedded in the flux rope. The flux cancellation implies the growth of the flux rope, making it broader and located higher with time, until it reaches an unstable position when too much flux is added \citep[e.g.][]{Aulanier2010}.

\begin{figure}[!t]
    \centering
    \includegraphics[width=\textwidth]{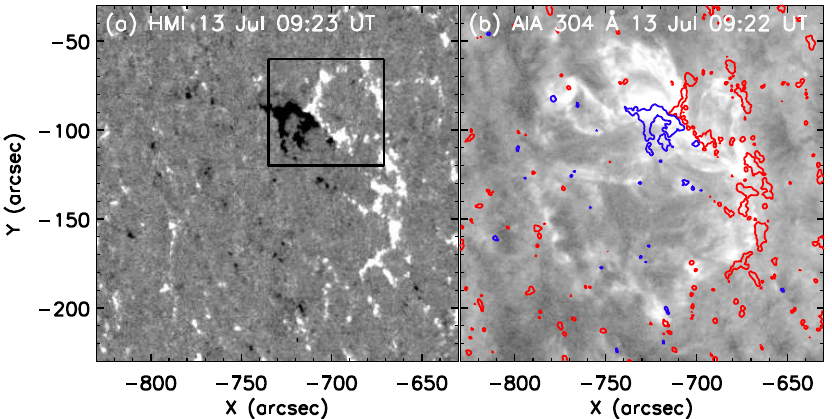}
    \caption{HMI magnetoram and AIA 304 \AA\ image at the time when the circular brightening was fully developed. (a) The black square in panel a surrounds the region that is shown in Figure \ref{fig:evolution_hmi}b. The magnetic field is saturated as in Figure \ref{fig:evolution_hmi}.
    (b) HMI positive and negative magnetic field contours of $\pm$ 50 G are over-plotted on the AIA image in red and blue colors. }
    \label{fig:circular_ribbon_flux}
\end{figure}

\begin{figure}[!t]
    \centering
    \includegraphics[width=\textwidth]{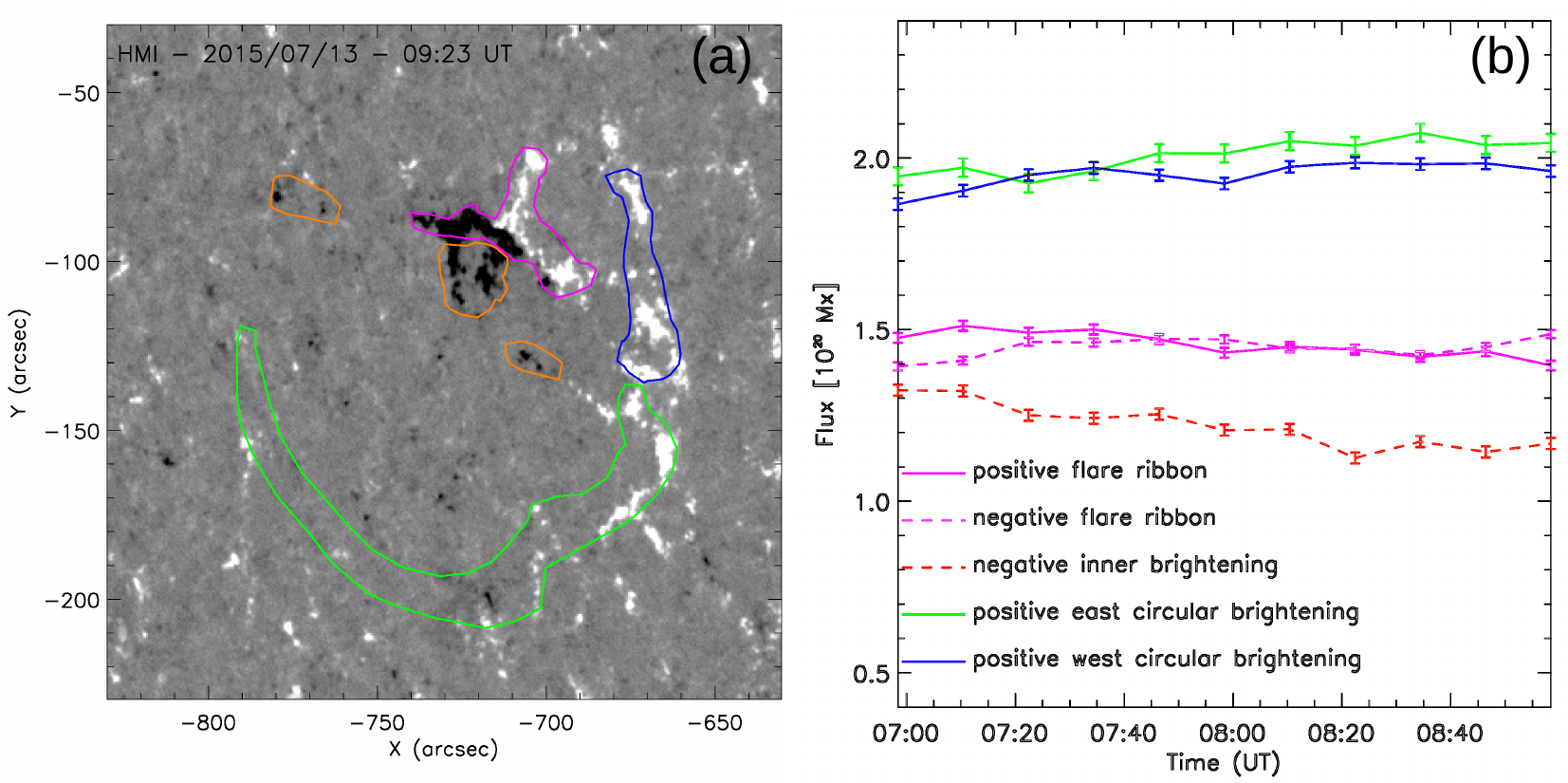}
    \caption{ 
    Magnetic flux in the flare ribbons and brightenings
    (a): The same HMI magnetogram as in panel a of Figure \ref{fig:circular_ribbon_flux} with over-plotted, using different colors, the locations of all the brightenings identified during the 13 July event at their maximum extension. 
    (b): Evolution of the magnetic flux within the contours shown in panel a computed from 7 UT 
until approximately the end of the pre-flare phase. The error bars in these curves are calculated in a similar way as those in Figure \ref{fig:flux}. }
    \label{fig:flux_all}
\end{figure}

An important information on the physics involved in this event is the amount of reconnected flux. As a first approach, this can be estimated computing the magnetic flux at the photospheric level encompassed by the flare ribbons and brightenings. 
 Figure \ref{fig:circular_ribbon_flux} displays an HMI magnetogram in panel a and an AIA 304 \AA\ image in panel b during the main phase of the B8.9 flare. The large circular brightening, which is the most extended one during this event, is located mostly southward of the erupting region, and extends both eastward and westward.   
As shown in panel b it is dominantly on the positive magnetic polarity in this anemone type configuration. Moreover, the negative polarity is fully surrounded by a positive polarity. 
Such circular-like ribbons in anemone type configurations have been reported and studied in several other examples (see references in Section \ref{sec:Introduction}).

Figure \ref{fig:flux_all}a shows the same magnetogram as in panel a of Figure \ref{fig:circular_ribbon_flux}. In this case we have over-plotted, using different colors, the locations of all the flare ribbons and brightenings at their maximum extension after overlaying the corresponding AIA 304 \AA\ images on the HMI magnetogram. We call these brightenings as positive flare ribbon, negative flare ribbon, negative inner brightening, eastern positive circular brightening, and western positive circular brightening. 

Figure \ref{fig:flux_all}b displays the evolution of the magnetic flux associated with these ribbons and brightenings from around 7:00 UT until around the end of the pre-flare phase. These curves show that all the computed fluxes remain roughly constant during this period of time, when no activity or emergence could perturb the magnetic field measurements.
The main uncertainties in these computations come from the imprecision in defining the boundaries of the brightenings. In particular, it is difficult to set the separation between the negative inner brightening and the negative flare ribbon. The same applies also to the separation between the positive east and the west circular brightenings. In Section \ref{sec:Scenario} we will discuss these flux measurements with the flux exchange occurring during magnetic reconnection in the context of a proposed scenario for the 13 July 2015 event. 

\section{Physical Processes Involved in the 13 July Event}
\label{sec:Physical_Processes}

In this section, we put the observations we have analyzed into context to understand the origin of the filament eruption, the two-ribbon flare, the presence of the circular ribbon, the EUV loop dynamics (expansion and contraction), and the CME characteristics. 

\begin{figure}
    \centering
    \includegraphics[width=\textwidth]{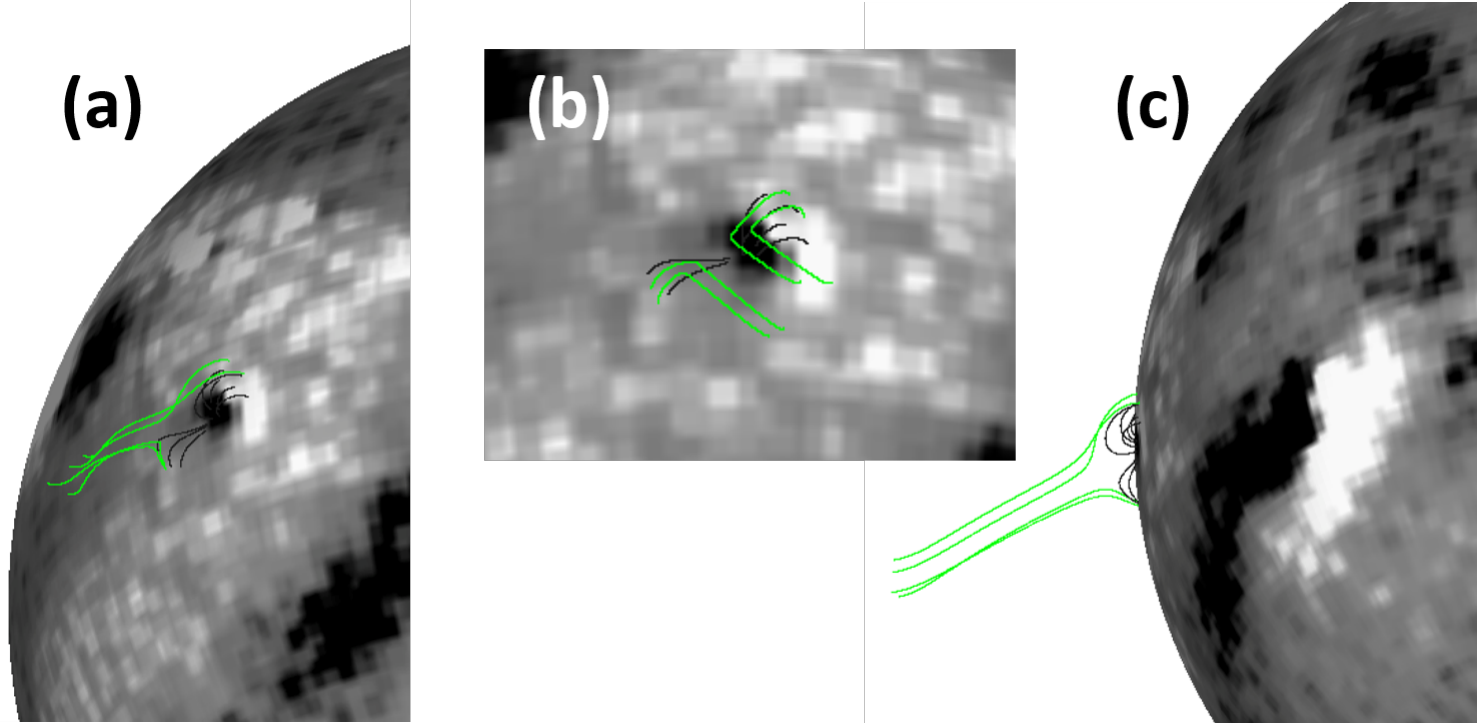}
    \caption{
     PFSS extrapolation based on the GONG-ADAPT synoptic map for 13 July at 08:00 UT, illustrating a fan-spine magnetic topology in the region of the filament eruption. Green lines indicate open magnetic field lines, while black lines trace closed loops. (a)  A magnetic null point is located between the open and closed field lines, delineating the topological configuration in its vicinity. In this panel, the center of the solar disk is located at $0^{\circ}$ of longitude and rotated by $-30^{\circ}$ in latitude.
    (b) Top view of the AR magnetic connectivity with the center of the solar disk rotated by $-30^{\circ}$ 
    in longitude and $-30^{\circ}$ in latitude. The configuration of the open field lines suggests a potential channeling path for the erupting filament. (c) Same as panel a, but the point of view is changed to have the null point at the eastern limb and at the same latitude. This viewpoint highlights the magnetic topology around the null point revealing two closed field regions below the fan surface and open field lines around the external spine. The estimated height of the null point is $\approx$ 47 Mm. The magnetic field values are saturated above/below $\pm$ 30 G in all panels.}

    \label{fig:pfss}
\end{figure}

\subsection{The Global Magnetic Field Topology}
\label{sec:PFSS}

The studied region has an excess of positive magnetic flux (Figures \ref{fig:flux_all} and \ref{fig:pfss}). This implies the existence of, at least, far-reaching connections, with scales comparable to the solar radius and most likely open magnetic field regions since the positive flux region is broad (Figure \ref{fig:pfss}a).  Then, a global solar magnetic field extrapolation is more suited for our region than a local one. 

To visualize the large-scale coronal connectivity in our analyzed region, we use a global potential-field source-surface (PFSS) extrapolation. The boundary condition of the model is the GONG Air Force Data Assimilative Photospheric Flux Transport (ADAPT) map on 13 July at 08:00 UT \citep[see][]{Worden2000}. The model is based on the Finite Difference Iterative Potential-Field Solver (FDIPS) code \citep{Toth11}. This code, which is freely available from the Center for Space Environment Modeling (CSEM) at the University of Michigan (\url{csem.engin.umich.edu/tools/FDIPS}), solves the Laplace equation for the magnetic field using an iterative finite-difference method. The spatial resolution for the model is $1^{\circ}$ in longitude (360 longitudinal grid points), $1.1\times10^{-2}$ in the sine of latitude (180 latitudinal grid points) and 1$\times$10$^{-2}$\,$\mathrm{R}_{\odot}$ in the radial direction. Our model for this particular field configuration, a low negative flux polarity surrounded by a much extended positive flux region, assumes that the field becomes purely radial at a height set to the value $1.5\,\mathrm{R}_{\odot}$.

The large-scale configuration above the erupting filament exhibits the topology of a pseudo-streamer, as expected for a region with a dominant positive flux embedding a weaker negative flux.  Such configuration implies the presence of magnetic null point and associated fan separatrix, which delineate open and close magnetic field. We illustrate this topology by tracing selected field lines in the vicinity of the null point in Figure \ref{fig:pfss}a. Green field lines, rooted in positive photospheric field, are open. This opening is not due to the source surface (located at 1.5 R$_\odot$), but starts at a much lower height because of the  negative flux is much weaker than the surrounding positive one. Black field lines represent two different sets of closed field lines situated below the fan dome. Green field lines surround the external part of the spine (starting at the null point). Panel b in Figure \ref{fig:pfss} shows the magnetic connectivity above the null point as viewed from the source surface height, with the disk center rotated by 
$-30^{\circ}$ in longitude and $-30^{\circ}$ in latitude.

Next, we estimate the height of the null point using the 3D visualization facilities of the PFSS code.
To minimize projection effects in our height estimation, we rotate the viewing angle by $55^{\circ}$ to the East in longitude, 
The estimated height of the null point is $\approx$ 47 Mm. 
This value is lower than the height of 56 Mm derived from the observations, as described in Section \ref{sec:EUV_Loop} (Figure \ref{fig:loops}).  In fact,  a higher height is expected when the configuration has sheared magnetic arcades compared to a potential field; this happens because the magnetic pressure is enhanced in a sheared configuration, so the lower arcade is more extended.
Notably, the estimated height is comparable to the radius of the circular ribbon ($\sim 55$  Mm), which is consistent with previous studies \citep[e.g.][]{Mandrini2014}. 

The fan lines have their footpoints in the region where we see the circular brightening during the filament eruption (Figure \ref{fig:circular_ribbon_flux}). The filament eruption occurs below the fan, in its northern part, i.e. below the set of northern closed field lines in panel c of Figure \ref{fig:pfss}. 
We conclude that during its eruption, the filament interacts with the overlying field lines and reconnection happens at the null-point and in its neighborhood.  This is a so-called interchange reconnection between open and closed field lines \citep[see, e.g.,][and references therein]{Crooker2012}. During this process, particles are accelerated and then transported along field lines towards the chromosphere. The following energy deposition creates the circular brightening at the positive footpoints of the reconnected field lines.   
The reconnected closed field lines also have a negative anchorage. This is where the inner brightening is observed (Figure \ref{fig:flux_all}). 

\begin{figure}[!t]
    \centering
    \includegraphics[width=0.8\textwidth]{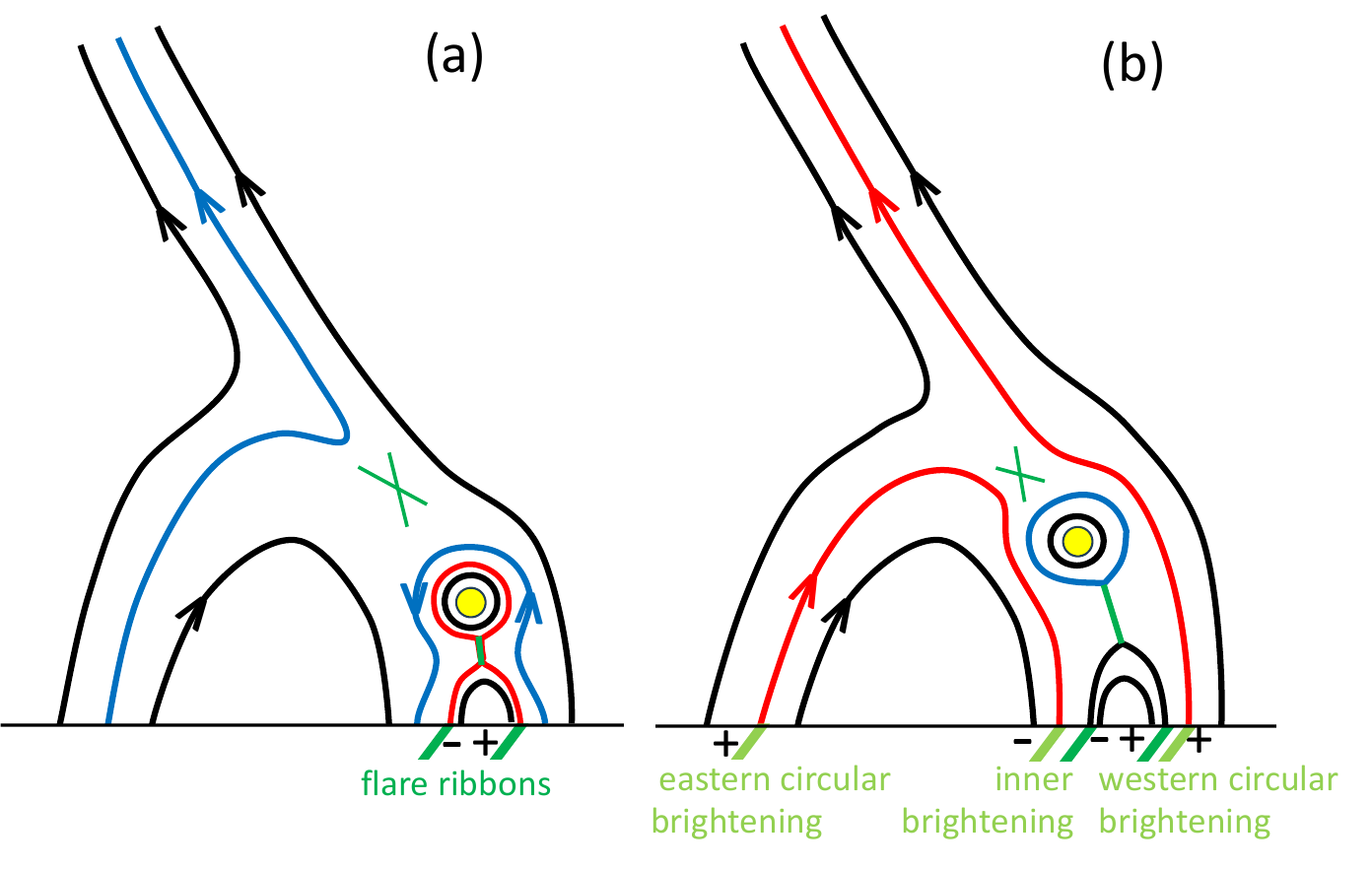}
    \caption{Schematic representation of the event including flare ribbons and the circular brightening. The sketch includes the two reconnection regions: one below the erupting filament, which results in  the standard two ribbons (panel a), and one above (at the null point) that implies the presence of the circular brightening and the inner brightening (panel b). The blue and red lines corresponds to field lines before and after reconnection, respectively. }
    \label{fig:cartoon}
\end{figure}

\subsection{A Scenario for the 13 July Event}
\label{sec:Scenario}

The studied filament is along a PIL where important magnetic flux cancellation is observed at the photospheric level (Figures \ref{fig:evolution_hmi} and \ref{fig:flux}).   From the sheared magnetic arcade present above the PIL, this flux cancellation induces magnetic reconnection at its footpoints. This builds a flux rope where dense and cold plasma could be caught, forming a stable filament.

As more flux is added up to the flux rope, the stabilizing effect of the downward magnetic tension of the overlying arcade decreases in strength and the flux rope grows in size and height, becoming unstable (Figure \ref{fig:kinematics}).  This induces further reconnection below the flux rope, most likely above the photosphere. This implies a peak in emission both in EUV and soft X-rays (Figure \ref{fig:goes_aia}).  In this case, later on, the magnetic configuration, traced by the filament, regains stability for about 10 min. 

Next, the flux rope becomes again unstable (Figure \ref{fig:kinematics}). This leads to the ejection of the filament and the B8.9 class flare (Figure \ref{fig:evolution}). The two J-shaped flare ribbons are typical of the eruption of a flux rope in a bipolar configuration \citep{Demoulin1996JGR}. A 2D sketch is shown in Figure \ref{fig:cartoon}a. The ejection of the flux rope induces strong reconnection below the flux rope, further building up the flux rope and injecting energy in the low atmosphere forming the flare loops and ribbons.

As the filament rises, Figure \ref{fig:kinematics}, its magnetic configuration interacts with the surrounding open magnetic field.  Indeed, the photospheric negative magnetic flux is surrounded by a much larger positive magnetic flux region (Figures \ref{fig:circular_ribbon_flux} and \ref{fig:pfss}a); so, the filament is embedded in a pseudo-streamer with a magnetic null point located above the closed magnetic field.  The ejected magnetic field induces a second reconnection process at this null point, which is expected to partially destroy the flux rope and also inject energy in the closed reconnected loops.  This evolution creates the two parts (east and west) of the circular brightening (circling the negative polarity)  and an inner brightening (Figures \ref{fig:circular_ribbon} and \ref{fig:flux_all}), as schematized in Figure \ref{fig:cartoon}b.  
The amount of photospheric magnetic flux in the east and west quasi-circular brightenings 
are comparable, $\sim 1.45 \times 10^{20}$ Mx, compared to $\sim 1.2 \times 10^{20}$ Mx for the inner brightening (Figure \ref{fig:flux_all}). In theory, the same reconnected flux should be involved in the three brightenings, and the above difference indicates the magnitude of the estimation error (which is larger than the statistical error shown in Figure \ref{fig:flux_all} because of the difficulty to determine precisely the magnetic flux associated to the brightenings).

Furthermore, coronal loop contraction followed by expansion are observed on the southern side of the filament eruption (Figures \ref{fig:loops} and \ref{fig:ts_loop}). This evolution is the consequence of magnetic reconnection at the null point. That is to say, first, new field connections are formed by reconnection above the existent loops and, as a consequence, they contract; second, because of the full ejection of the flux rope, the loops can expand upward and fill the remnant less magnetized region.

The flux rope ejection is detected as a CME in LASCO coronagraph (Figure \ref{fig:CME}). The bright and round-shape structure observed in the scattered light shows that part of the flux rope does survive. On the southern side,  plasma with a radial motion is observed. Its brightness indicates that it is denser than the background. This is most likely the heated plasma, from the closed loops and so dense, launched in the open field lines after the interchange reconnection process. In this regard, the expected fourth ``brightening'' of the null-point reconnection is not present in the low atmosphere but in the solar wind, since one extreme of the reconnected field lines is open.

 Finally, by combining a variety of observations we can justify the scenario drawn in Figure \ref{fig:cartoon} as summarized in Section \ref{sec:Conclusion}. 

\section{Summary and Conclusions} \label{sec:Conclusion}

In this article, we present the analysis of a filament eruption. The filament 
erupts on 13 July 2015 and is accompanied by a CME and a flare of GOES B8.9 class. Our results are summarized as follows: 
\begin{itemize}
   \item{ The studied filament lies on the central PIL of an evolving AR. The spatial and temporal evolution of the photospheric magnetic field shows that continuous magnetic flux cancellation occurs along the PIL. This is expected to drive magnetic reconnection beneath the filament, leading to the buildup of the flux rope, its gradual rise, up to its instability.}\\
    
    \item{The filament erupts in three different phases. First, a pre-flare enhancement is observed in AIA wavelengths and GOES data. The filament starts to rise during this pre-flare enhancement and it reaches a height of $\approx$ 22 Mm but does not erupt. Second, the filament rises slowly with a speed of 6 -- 9 \kms\ likely as a consequence of magnetic field cancellation at the PIL which further builds up the flux rope embedding the filament. Third, about 20 min after the pre-flare enhancement, a B8.9 class flare starts. Then, the filament accelerates upward and erupts with a speed up to 196 \kms. The observed plasma traces an untwisting motion. From the observed flare ribbons in AIA images (inverse J-shape), we conclude that the twist in the flux rope is left handed.}\\
    
    \item {A remarkable feature of this event is the reconnection between the erupting flux rope and overlying field lines, which gives rise to a circular brightening on the southern side of the erupting filament region. A PFSS extrapolation reveals a fan-spine topology associated to a magnetic null point located above the erupting filament. The fan field lines are rooted at the circular brightening. From AIA observations and the PFSS model the null point is located at a height between 47 -- 56 Mm above the photosphere. }\\

    \item {The EUV loops located on the southern side of the filament, so in another connectivity cell,  contract after the rise of filament with a speed in the range of 4 -- 24 \kms. This is followed by expansion with a lower speed of 4 -- 5 \kms. The first stage is due to accumulation of overlying reconnected flux and the second stage to the coronal ejection of the flux rope which leaves a magnetic void to be filled out by the loop magnetic field.}\\

    \item {A CME, associated with the filament eruption, has a speed of $\approx$ 550 \kms, decelerates with $\approx$ 14 m s$^{-2}$, and has a total projected angular width of 51$\degree$. The CME shows a classical flux rope ejection and also plasma ejected along open field lines located south of this flux rope.}   
\end{itemize}

The variety of observations we have analyzed, combined with a global coronal magnetic model, supports the scenario we draw in Section \ref{sec:Physical_Processes} (see Figure \ref{fig:cartoon}) and leads us to conclude on the role and outcome of each of the stages exemplified by this scheme. The scenario includes many steps that have been observed, at least partially, in previous filament eruption examples.   Magnetic flux cancellation at the PIL is confirmed to be a key mechanism to form a flux rope and later bring it to instability. The ejection is facilitated by the surrounding open magnetic field and the presence of a null point above the erupting flux rope. Magnetic reconnection at the null-point location induces the presence of extra ribbon-like brightenings, which results in a complex emission distribution. This distribution outlines the magnetic field topology  \citep[separatrices and, more generally, quasi-separatrix layers, QSLs,][]{Demoulin1996,Demoulin1997}. Finally, the reconnection process at the null point is not sufficient to process the full flux rope because a trace of it is present in coronagraph observations, together with a radial dense plasma region as evidence of interchange reconnection between closed and open field lines at the null point.  

In conclusion, the studied eruption presents non-classical features, as follows. During the eruption, the erupting magnetic field reconnects with the overlying open magnetic field resulting in the formation of a circular ribbon encircling the eruption site (where the classical two ribbons are present). A potential-field source-surface (PFSS) extrapolation indicates a fan-spine magnetic configuration associated to a magnetic null point. A notable aspect of this event is the change in the dynamics of neighboring coronal loops, which contract during the pre-flare phase and subsequently expand — some even rising above their initial positions. Our analysis provides an explanation of the mechanisms driving these phenomena and sheds light on the interaction between the erupting filament magnetic field and its surrounding coronal structures.
    
\begin{acks}
The author thank the reviewer for the valuable comments and suggestions on the manuscript. We thank the open data policy of NASA's SDO mission. We also acknowledge the SOHO and GONG instruments for providing the archival data. The author thank Dr. Hebe Cremades for the discussion on the CME analysis in the present study.
\end{acks}

\appendix   
\section{Data Sets }
\label{sec:obs}

For this study, we used data from:
\begin{enumerate}
\item{\bf Solar Dynamics Observatory (SDO):}
The SDO \citep{Pesnell2012} is a space mission that obtains high spatio-temporal resolution data of the solar full disk.  We analyze the data from two of its instruments, which are the Atmospheric Imaging Assembly \citep[AIA;][]{Lemen2012} and the Helioseismic and Magnetic Imager \citep[HMI;][]{Scherrer2012}. AIA gives us the opportunity to observe various layers of the solar atmosphere in different wavelengths. It takes data with seven EUV (94, 131, 171, 193, 211, 304, and 335 \AA), two UV (1600 and 1700 \AA), and one white light (4500 \AA) filters. The temporal and pixel resolution AIA is 12 sec and 0.6$''$, respectively. For the current study, we have used data in 131, 171, 193, and 304 \AA\ wavelengths to observe different aspects of the eruption. HMI observes the full Sun with a temporal resolution of 45 sec and a pixel resolution of 0.5$''$. It provides details of the photospheric magnetic field using the 6173 \AA~line and of sunspots using continuum images (4500 \AA). We used the data of HMI to analyze the variations of photospheric magnetic field before and during the eruption. \\

\item{\bf Global Oscillation Network Group (GONG):}
The GONG \citep{Harvey2011} observes the full Sun in H$\alpha$ filtergrams. The temporal and spatial resolutions are 1 min and 2$''$, respectively. We use data from GONG,  particularly from El Teide Observatory located on Tenerife, Canary Islands, Spain,
to see the spatial evolution of the filament eruption and associated flare in the solar chromosphere. 
\end{enumerate}

All data obtained from the instruments mentioned above are processed using SolarSoft \citep[SSW,][]{Freeland1998}. SSW is a package for solar image processing written in Interactive Data Language (IDL).
 
\begin{authorcontribution}
P. Devi did the data analysis and wrote a substantial part of the manuscript. C.H.M., R.C., and P. D\'emoulin wrote a significant part of the manuscript and contributed to the event interpretation. G.D.C. did the computation of the magnetic flux. C.M.C and G.D.L. did the potential-field source surface (PFSS) extrapolation which supported the event interpretation. All the authors reviewed the manuscript.   
\end{authorcontribution}

\begin{fundinginformation}
R.C. acknowledges the support from DST/SERB project No. EEQ/2023/000214. The authors C.H.M. and G.D.C. are members of the Carrera del Investigador Cient\'{\i}fico of the Consejo Nacional de Investigaciones Cient\'{\i}ficas y T\'ecnicas (CONICET) of Argentina. C.H.M. and G.D.C. acknowledge financial support from the Argentinian grants PICT 2020-03214 (ANPCyT) and PIP 11220200100985 (CONICET).
P.D. acknowledges the support by the Centres of Excellence scheme of the Research Council of Norway, project number 262622.
\end{fundinginformation}

\begin{dataavailability}
The data sets analyzed during the current study are available at \url{http://jsoc.stanford.edu/}, \url{https://cdaw.gsfc.nasa.gov/}, and \url{https://gong2.nso.edu/archive/patch.pl?menutype=s}.
\end{dataavailability}



\begin{ethics}
\begin{conflict}
The author C.H.M. is Editor-in-Chief of the journal Solar Physics; the article underwent a standard single-blind peer review process. The authors declare that they have no other conflicts of interest.
\end{conflict}
\end{ethics}


\IfFileExists{\jobname.bbl}{} {\typeout{}
\typeout{****************************************************}
\typeout{****************************************************}
\typeout{** Please run "bibtex \jobname" to obtain} \typeout{**
the bibliography and then re-run LaTeX} \typeout{** twice to fix
the references !}
\typeout{****************************************************}
\typeout{****************************************************}
\typeout{}}

\bibliographystyle{spr-mp-sola}
\bibliography{references}


\end{document}